\documentclass[fleqn]{article}

\usepackage{amsmath,amssymb}
\usepackage{graphicx}

\def\tref#1#2#3{{#1}~(#2)~#3}

\begin{document}

\title{{\sf  Pad\'e Estimate of QCD's Infrared Boundary}}
\author{F.A.~Chishtie\thanks{email: fachisht@uwo.ca}, V.~Elias\thanks{email: velias@uwo.ca}\\
  \textsl{Department of Applied Mathematics}\\
  \textsl{University of Western Ontario}\\
  \textsl{London, Ontario, N6A 3K7}
  \textsl{Canada}\\[5pt]
  T.G.~Steele\thanks{email: Tom.Steele@usask.ca}\\
  \textsl{Department of Physics and Engineering Physics}\\
  \textsl{University of Saskatchewan}\\
  \textsl{Saskatoon, Saskatchewan, S7N 5E2}\\
  \textsl{Canada}
 }
\maketitle
\begin{abstract}
A mass scale $\mu_c$ representing the boundary between effective theories of strong interactions and 
perturbative QCD (infrared and ultraviolet regimes) exists when the $\beta$-function is 
characterized by a simple pole.  This behaviour, which is known to occur in $N=1$ supersymmetric Yang-Mills theories, 
leads to an infrared attractor in the evolution of the coupling constant, with the mass scale $\mu_c$ of 
the attractor providing a natural boundary between the infrared and ultraviolet regimes.  
It is demonstrated that $[2|2]$, $[3|1]$ and $[1|3]$ Pad\'e-approximant versions of the  
three-flavour ($n_f=3$) QCD $\beta$-function 
each contain a simple pole corresponding to such an infrared attractor.  
All three approximants, separately considered, are 
seen to lead to nearly equivalent estimates for the mass scale $\mu_c$.
\end{abstract}
Although QCD is well-understood both qualitatively and quantitatively as a perturbative theory for
the strong interactions, we have a surprisingly limited amount of information about the infrared
boundary of its perturbative domain. Such a boundary (which is also anticipated by arguments in
which hadrons and quarks are dual field variables for weak and strong QCD \cite{Seiberg}) must exist to
separate effective strong interaction theories ({\it e.g.,} chiral perturbation theory, linear sigma model,
{\it etc.}) from the higher-momentum region characterised by the perturbative quantum field theory of
quarks and gluons.  Consequently, one can argue that the infrared behaviour of the perturbative QCD
coupling should {\em not} be characterised by smooth evolution to an infrared-stable fixed point, but rather
by behaviour that would clearly separate the infrared domain of effective strong-interaction theories 
from perturbative QCD. Indeed, there exists both lattice \cite{Iwasaki}, analytical \cite{analytical}, 
and Pad\'e-approximant \cite{Elias,Ndili} corroboration for the idea that an infrared-stable fixed point 
does not occur within QCD unless
the number of active fermion flavours contributing to the evolution of the QCD couplant is
substantially larger than three.

\begin{figure}[hbt]
\centering
\includegraphics[scale=0.75]{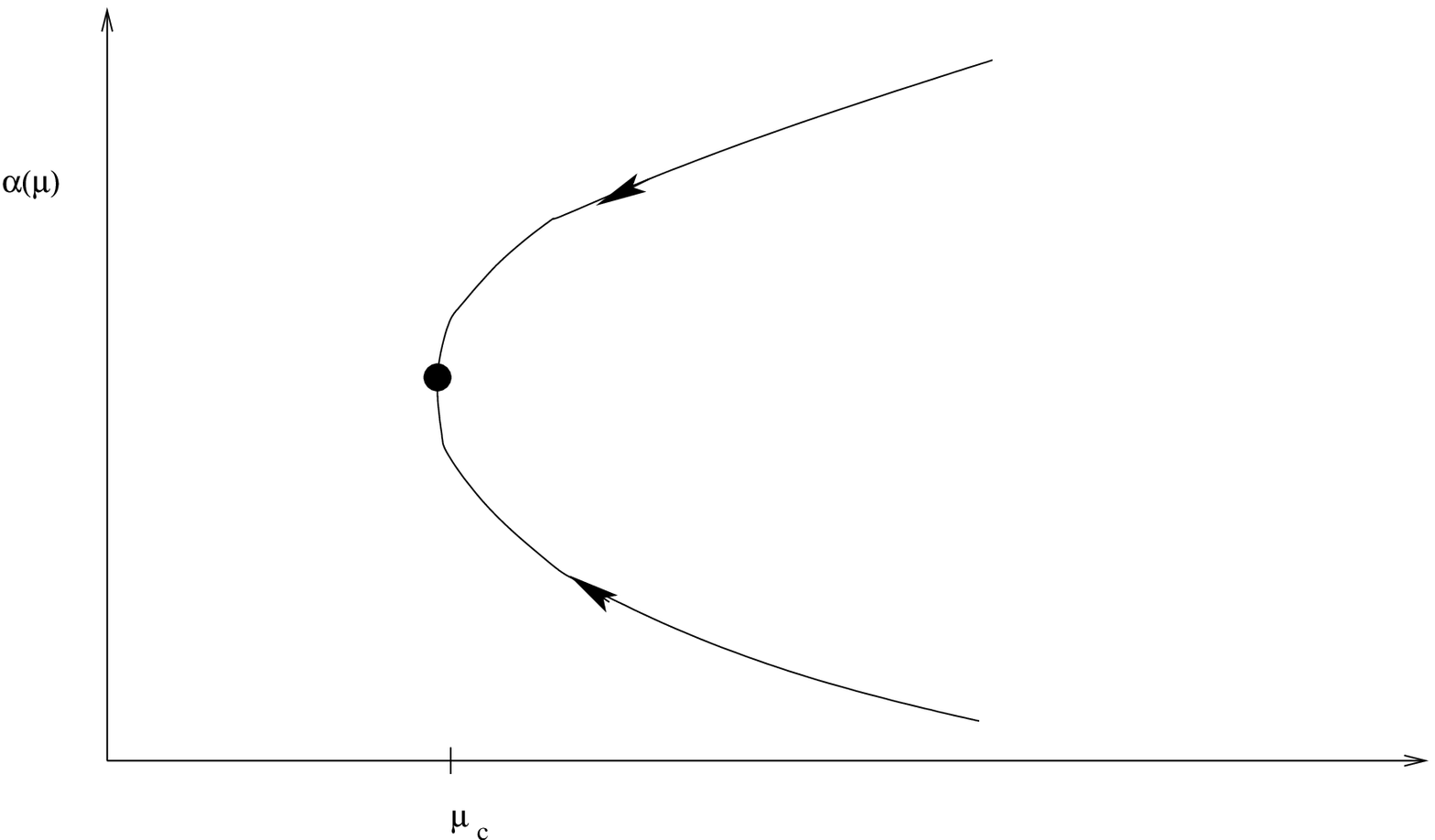}
\caption{Qualitative behaviour of the running coupling $x\left(\mu\right)$ in an asymptotically-free
theory with an infrared attractor devolving from a simple pole in the $\beta$-function.
}
\label{ir_attractor_fig}
\end{figure}

A clear demarcation between the infrared and ultraviolet regions would occur if the $n_f=3$ QCD  
$\beta$-function were characterised not by a positive zero corresponding to an infrared-stable fixed point,
but rather by a positive {\em pole}, an infrared-attractive singularity in the  $\beta$-function occurring at a
momentum scale which necessarily corresponds to a lower bound on the domain of the running QCD
coupling constant (assumed here to be real).  Such behaviour, schematically presented in 
Fig.~\ref{ir_attractor_fig}, is
known to characterise the exact  $\beta$-function for $N=1$ supersymmetric Yang-Mills theory in the NSVZ
renormalisation scheme \cite{NSVZ}, a ``supersymmetric gluodynamics'' whose  $\beta$-function pole forms an 
infrared-attractive terminal point for coupling-constant evolution within the theory's 
asymptotically-free phase \cite{Kogan}. There exists evidence from Pad\'e-approximant methods that 
similar dynamics may
also characterise the infrared region of QCD \cite{Elias}. In the present note, we discuss whether such
methods can provide information as to the actual mass scale which separates the infrared and
ultraviolet domains of strong interaction physics, as well as the strong-interaction couplant
magnitude characterising this mass scale.

Weighted asymptotic Pad\'e-approximant (WAPAP) methods have been utilised to estimate the 
$n_f$-dependence of the five-loop contribution ($\beta_4$) to the QCD  $\beta$-function, 
defined here for $x(\mu)\equiv\alpha_s(\mu)/\pi$ 
to be
\begin{equation}
\mu^2\frac{\mathrm{d}x}{\mathrm{d}\mu^2}=-\sum_{k=0}^\infty \beta_k x^{k+2}=-\beta_0x^2
\left(1+R_1x+R_2x^2+R_3x^3+R_4x^4+\ldots\right)\quad .
\label{beta_def}
\end{equation}               
For $n_f = 3$, the known $\overline{{\rm MS}}$ $\beta$-function coefficients in (\ref{beta_def}) are \cite{Ritbergen}
\begin{equation}
\beta_0=\frac{9}{4}\quad ,\quad \beta_1=4\quad ,\quad \beta_2=\frac{3863}{384}\quad ,\quad
\beta_3=47.2280\quad .
\label{beta_coeffs}
\end{equation}
The WAPAP estimate of the $n_f$-dependence of the five-loop term (inclusive of quadratic-Casimir
contributions to $\beta_3$) is found to be \cite{Ellis}
\begin{gather}
\beta_4\cong\frac{1}{4^5}
\left( 7.59\times 10^{5}-2.19\times 10^5 n_f+2.05\times 10^4 n_f^2-49.8n_f^3-1.84n_f^4\right)
\stackrel{n_f=3}{\longrightarrow} 
278\quad ,
\label{beta4_value}\\
R_4=\frac{\beta_4}{\beta_0} 
\stackrel{n_f=3}{\longrightarrow}
124\quad .
\label{R4_value}
\end{gather}
This estimate may be used to construct non-trivial  $N + M = 4$ $[N\vert M]$-Pad\'e approximants that reproduce the
known coefficients $R_1\mbox{--}R_3$ and the estimate (\ref{R4_value}) for 
$R_4$ within the  $\beta$-function series of (\ref{beta_def}):
\begin{gather}
\beta^{[2\vert 2]}(x)=-\frac{9}{4}x^2\frac{\left[1-5.4983x-1.9720x^2\right]}{\left[1-7.2762x+6.4923x^2\right]}
\label{beta_22}\\[5pt]
\beta^{[1\vert 3]}(x)=-\frac{9}{4}x^2\frac{\left[1-5.7397x\right]}{\left[1-7.5174x+8.8933x^2-3.1896x^3\right]}
\label{beta_13}\\[5pt]
\beta^{[3\vert 1]}(x)=-\frac{9}{4}x^2\frac{\left[1-4.1155x-6.0058x^2-5.3589x^3\right]}{\left[1-5.8932x\right]}
\label{beta_31}
\end{gather}
All three approximants above exhibit a positive pole that precedes any positive zeroes, consistent
with infrared dynamics analogous to those of NSVZ supersymmetric gluodynamics, as discussed
above.\footnote{The fact that a positive pole is always found to precede any  
positive zeroes for all three approximants has already been established \protect\cite{Elias} for {\em arbitrary} values of $R_4$.}
This first positive pole, which corresponds to the couplant magnitude at the 
infrared-boundary momentum scale $\mu_c$, is surprisingly comparable for all three approximants:
\begin{gather}
[2|2]:\quad x\left(\mu_c\right)=0.160\quad,
\nonumber\\
[1|3]:\quad x\left(\mu_c\right)=0.162\quad,
\label{beta_pole}\\
[3|1]:\quad x\left(\mu_c\right)=0.170\quad .
\nonumber
\end{gather}
The fact that {\em all three} approximants exhibit a positive pole of nearly equivalent magnitude (which
precedes any positive zeroes)
is indicative that such a pole is not likely to be an artefact defect pole \cite{Baker},
but rather a manifestation of a true pole within the underlying all-orders  $\beta$-function.  
It is to be
noted that the pole couplant magnitude, as estimated in (\ref{beta_pole}),  is itself sufficiently small for
perturbative physics to remain viable near the infrared boundary. This  behaviour
is very different from an infrared-slavery scenario in which the QCD couplant grows non-perturbatively large in the
vicinity of a deep-infrared Landau singularity. 

Similar consistency of poles obtained from differing Pad\'e approximants to the perturbative  
$\beta$-function has already been shown to characterise supersymmetric gluodynamics in both NSVZ and
in DRED renormalization schemes \cite{EliasJPG}.  We reiterate that the pole in the former of these two
schemes is known to occur from the all-orders  $\beta$-function expression, whether derived via NSVZ
instanton calculus \cite{NSVZ} or via imposition of the Adler-Bardeen theorem upon the supermultiplet
structure of the theory \cite{EliasJPG,Jones}.

One can utilise the infrared-attractive couplant values (\ref{beta_pole}) in order to obtain 
separate estimates of the
infrared boundary $\mu_c$ for each approximant considered. For specific
$n_f=3$ $[N\vert M]$ approximant versions of
the QCD  $\beta$-function, one finds that
\begin{equation}
\mu_c^{[N|M]}=m_\tau\exp\left[
\frac{1}{2}\int\limits_{x\left(m_\tau\right)}^{x\left(\mu_c\right)}\,\frac{\mathrm{d}x}{\beta^{[N|M]}(x)}
\right]\quad .
\label{mu_c_expr}
\end{equation}
We utilise the approximants (\ref{beta_22})--(\ref{beta_31}) within the integrand
of (\ref{mu_c_expr}), as well as the respective values (\ref{beta_pole}) for $[2| 2]$,  $[1| 3]$, and  $[3| 1]$ 
approximant values of $x\left(\mu_c\right)$ for the upper bound of
integration.  
For the lower bound of integration, we assume
$\alpha_s^{\overline{MS}}\left(m_\tau\right)=\pi
x\left(m_\tau\right)=0.33\pm 0.02$, consistent with recent analyses \cite{aleph}.
We then obtain via (\ref{mu_c_expr}) the following values for the infrared-boundary momentum scale:
\begin{gather}
\mu_c^{[2|2]}=1.14\pm 0.11\,{\rm GeV}
\nonumber\\
\mu_c^{[1|3]}=1.13\pm 0.11\,{\rm GeV}
\label{mu_c_values}\\
\mu_c^{[3|1]}=1.09\pm 0.11\,{\rm GeV}
\nonumber
\end{gather}
These results not only exhibit remarkable consistency with each other, but also support the identification
of perturbative QCD's infrared boundary with a momentum scale at (or somewhat above) the mass
scale characterising nucleons and the vector meson octet. This picture  is quite different from the
usual one of an ${\cal O}(300 \,{\rm MeV})$ value for the Landau singularity  
$\Lambda_{QCD}$ characterising coupling constant
evolution via the {\em truncated}  $\beta$-function series. We reiterate that the estimates (\ref{beta_pole}) 
and (\ref{mu_c_values}) for the
infrared terminus of the couplant evolution within the asymptotically free phase of $n_f = 3$ QCD are
consistent with both the applicability of controllably-perturbative QCD down to ${\cal O}(1 \,{\rm GeV})$
momentum scales, as well as  the necessity for alternative descriptions ({\it e.g.} effective Lagrangians and 
hadronic field variables) to characterise sub-GeV 
(or sub-$4\pi f_\pi$ \cite{manohar})
strong-interaction physics.

The results (\ref{mu_c_expr}) and (\ref{mu_c_values}) are, of course, sensitive to input information. 
An across-the-board $150\,{\rm MeV}$ decrease from (\ref{mu_c_values}) in the estimated range of 
$\mu_c$ is seen to occur if we
choose the three-flavour threshold at $\mu_t=m_c\left(m_c\right)\cong 1.25\,{\rm
GeV}$ \cite{PDG}. More significantly, 
the set
of values (\ref{mu_c_values}) for $\mu_c$ relies ultimately on ref.\  \cite{Ellis}'s WAPAP estimate 
(\ref{beta4_value},\ref{R4_value}) for the five-loop contribution
to the QCD  $\beta$-function.
The $n_f = 3$ estimate obtained via (\ref{beta4_value}) involves relatively small differences of large numbers.
Consequently, the individual coefficients of powers of $n_f$ are likely to be more accurate than the
estimate of  $\beta_4$ and $R_4$ obtained by evaluating the full polynomial in $n_f$.\footnote{ 
Such is the case for a prior asymptotic Pad\'e-approximant prediction of the four-loop 
term \protect\cite{Karliner},
$\beta_3^{pred} = \{23600 - 6400n_f + 350n_f^2 +1.499n_f^3\}/256$, 
whose polynomial coefficients are in good
term-by-term agreement with those of the subsequent exact calculation \protect\cite{Ritbergen},  
$\beta_3^{true} = \{29243.0 -
6946.30n_f + 405.089n_f^2 + 1.499n_f^3\}/256$. When $n_f =3$, however, the estimated value 
$\beta_3^{pred} = 29.65$
differs from  $\beta_3^{true} = 47.228$ by a relative-error (37\%) whose magnitude is a 
factor of two or more
larger than that of the relative error characterising each estimated polynomial coefficient (19\%,
7.9\%, and 14\%, respectively).}
 
     To test the uniformity of the infrared-boundary mass scale obtained via different
approximants, let us choose $R_4$ for each approximant so as to ensure the occurrence of a positive pole
at  $\alpha_s\left(\mu_c\right) =  \pi/4$ ({\it i.e.,} at $x = 1/4$). 
This particular choice is motivated both as the critical value of the
strong coupling for chiral symmetry breaking \cite{fomin} as well as by the couplant value characterising
Nambu-Jona-Lasinio and linear-sigma-model approaches toward the generation of a dynamical quark
mass \cite{sigma}.

The appropriate set of approximants for the $n_f=3$ $\beta$-function series
\begin{equation}
\mu^2\frac{\mathrm{d}x}{\mathrm{d}\mu^2}=-\frac{9}{4}x^2\left[
1+\frac{16}{9}x+\frac{3863}{864}x^2+20.9902x^3+R_4x^4+\ldots
\right]
\label{three_flav_beta}
\end{equation}
with $R_4$ arbitrary is
\begin{gather}
\beta^{[2|2]}(x)=-\frac{9}{4}x^2
\frac{\left[1+\left(7.19456-0.102610R_4\right)x+\left(-11.3292+0.0756438R_4\right)x^2\right]}{\left[1+
\left(5.41678-0.102610R_4\right)x+\left(-25.4301+0.258062R_4\right)x^2\right]}~ ,
\label{beta_22_R4}\\
\beta^{[1|3]}(x)=-\frac{9}{4}x^2
\frac{\left[1+\left(5.80845-0.0933552R_4\right)x\right]}{\left[1+\left(4.03067-0.0933552R_4\right)x
+\left(-11.6367+0.165965R_4\right)x^2+\left(-18.3242+0.122349R_4\right)x^3\right]}~ ,
\label{beta_13_R4}\\
\beta^{[3|1]}(x)=-\frac{9}{4}x^2
\frac{\left[1+\left(1.77778-0.0476412R_4\right)x+\left(4.447106-0.0846954R_4\right)x^2
+\left(20.9902-0.213007R_4\right)x^3\right]}{\left[1-0.0476412R_4x\right]}~.
\label{beta_31_R4}
\end{gather}
The first positive pole for all three approximants is seen to precede any positive zeroes and is clearly
dependent on the value of $R_4$.  
One can easily show that all three approximants acquire a positive pole at $x = 1/4$ for very similar
values of $R_4$:
\begin{gather}
[2|2]:\quad R_4=80.307
\nonumber\\
[1|3]:\quad R_4=89.925
\label{critical_R4}\\
[3|1]:\quad R_4=83.961
\nonumber
\end{gather}
Given  $\alpha_s\left(m_\tau\right)$ and  $x\left(\mu_c\right) = 1/4$, as ensured by these respective 
choices for $R_4$, one can utilise (\ref{mu_c_expr}) to
estimate corresponding mass scales $\mu_c$  for the three approximants. In Table \ref{mu_tab}, such values are
obtained via couplant evolution from three different choices for
$\alpha_s\left(m_\tau\right)$ in the range
$\alpha_s\left(m_\tau\right)=0.33\pm 0.02$ \cite{aleph}.
 In all these estimates,     three-flavour coupling constant evolution is
assumed to be valid below $m_\tau$.

Table \ref{mu_tab} shows striking uniformity in the values for the infrared boundary mass scale $\mu_c$
obtained via three distinct Pad\'e-approximants.   Moreover, the three approximants generate values
for $\mu_c$ very near the  $\rho$-meson mass when the  lowest value for  $\alpha_s\left(m_\tau\right)$ 
is chosen. 

\begin{table}
\centering
\begin{tabular}{||l|c|c|c||}
\hline\hline
   & $\alpha_s\left(m_\tau\right)=0.31$ 
& $\alpha_s\left(m_\tau\right)=0.33$ & $\alpha_s\left(m_\tau\right)=0.35$ 
\\
\hline\hline
$\mu_c^{[2|2]}$ & 785~MeV & 874~MeV & 960~MeV   
\\
\hline
$\mu_c^{[1|3]}$ & 778~MeV & 867~MeV & 952~MeV  
 \\
\hline
$\mu_c^{[3|1]}$ & 775~MeV & 863~MeV & 948~MeV 
  \\
\hline\hline
\end{tabular}
\caption{Values of $\mu_c^{[N|M]}$, as obtained via (\ref{mu_c_expr}),  for $[N| M]$ 
approximants to the
$n_f = 3$ QCD  $\beta$-function. The (unknown) five-loop contribution 
$R_4$ to the perturbative series (\protect\ref{three_flav_beta})
is chosen (see (\protect\ref{critical_R4})) to ensure a 
pole at $x\left(\mu_c\right) = 1/4$ for all three Pad\'e-approximant cases.}
\label{mu_tab}
\end{table}

\begin{figure}[hbt]
\centering
\includegraphics[scale=0.5,angle=270]{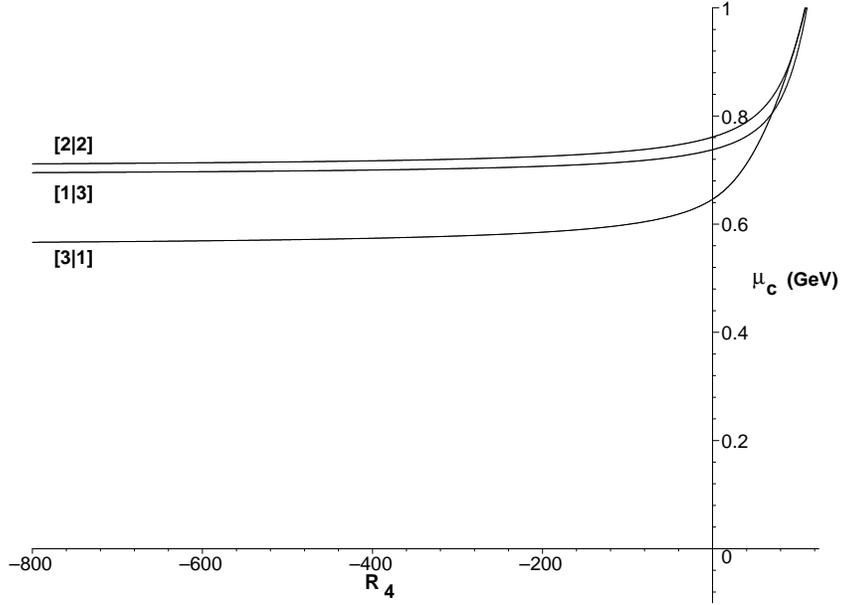}
\caption{Dependence of the infrared-boundary scale $\mu_c$ on the five-loop $\beta$-function coefficient 
$R_4=\beta_4/\beta_0$, based upon $n_f=3$ evolution of the couplant from an initial value of
$\alpha_s\left(m_\tau\right)=0.33$.  We only plot values 
of $R_4$ less than the WAPAP estimate (\ref{R4_value}); if $R_4$ is allowed to increase much past this value, the first positive poles of (\ref{beta_22_R4}--\ref{beta_31_R4}) are soon seen to be reached at values of $\mu$ exceeding the four-flavour 
threshold $\mu_t$.
}
\label{mu_c_fig}
\end{figure}

     Indeed, Figure \ref{mu_c_fig} demonstrates that the infrared boundary for all three approximants is nearly
equivalent {\em even if} $R_4$ {\em is allowed to  vary arbitrarily}.
The corresponding mass scale $\mu_c$, as plotted in Figure \ref{mu_c_fig}, is
obtained for each approximant by substituting (\ref{beta_22_R4}--\ref{beta_31_R4}) into the integrand of 
(\ref{mu_c_expr}), with (\ref{mu_c_expr})'s upper bound of integration identified with the first positive poles
of (\ref{beta_22_R4}--\ref{beta_31_R4}).\footnote{When $R_4$ becomes negative, the $[3|1]$ approximant 
version of the $n_f=3$ $\beta$-function no longer exhibits a positive pole (or any positive zeroes, as would be 
the case with an infrared stable fixed point), but still exhibits a Landau singularity extractable from 
(\ref{mu_c_expr}) provided the upper bound of integration in (\ref{mu_c_expr}) is replaced by infinity.  
The $[3|1]$ approximant curve in Figure \protect\ref{mu_c_fig} displays the mass scale for this Landau 
singularity when $R_4$ is negative. The figure shows this mass scale to be consistent with the mass scales 
characterising the  simple poles within the $[2|2]$- and $[1|3]$-approximant versions of the $\beta$-function for the same negative values of $R_4$.}
For a given choice of $R_4$, Figure \ref{mu_c_fig} shows that the infrared-boundary mass scales characterising 
all three approximants are surprisingly close.  
 The ${\cal O}(600 \,{\rm MeV})$ 
lower bound evident from the figure
for all three approximants is particularly striking. Different Pad\'e
approximants to the $n_f=3$ QCD  $\beta$-function thus appear to be quite
 consistent in the infrared dynamics they
predict, suggestive that similar pole-driven dynamics may characterise the infrared boundary of QCD itself.  

The above results have all been obtained in the $\overline{{\rm MS}}$
scheme,  an explicitly gauge-invariant renormalization procedure for which
the $\beta$-function has been explicitly calculated to four-loop order,
thereby facilitating our use of Pad\'e approximants to the series
(\ref{beta_def}).  Moreover, perturbative QCD phenomenology is more
readily available (and more likely to be corroborated) in  $\overline{{\rm
MS}}$ than in other renormalization schemes.  The question remains,
however, as to whether the infrared boundary we observe is a peculiarity
of the  $\overline{{\rm MS}}$ scheme we employ.   In eq.\ (\ref{beta_def}),
the coefficients $R_2,~R_3,~R_4,~\ldots$, (corresponding to $\beta_k$ with
$k\ge 2$) are all scheme-dependent, and therefore negotiable within the
context of formal perturbative QCD.  Indeed, a QCD renormalization scheme
proposed by 't Hooft \cite{thooft} 
in which $\beta_k=0$ for $k\ge 2$
is guaranteed {\em by construction} to
be free of $\beta$-function poles.\footnote{Curiously, the existence and approximate
size of the $n_f$ threshold for $\beta$-function {\em zeroes} ({\it i.e.,}
infrared-stable fixed points) in the 't Hooft scheme is
corroborated by Pad\'e-approximants constructed from the known terms of
the  $\overline{{\rm MS}}$ $\beta$-function \protect\cite{Elias}. }

Recent work \cite{cvetic} has developed a procedure by which the leading
renormalization-scheme dependence ({\it i.e.,} explicit dependence on
$R_2$) can be eliminated from a perturbative QCD result.  This approach
leads ultimately to $\beta$-functions characterized by the remaining
renormalization-scheme parameters ({\it i.e.,} $R_3,~\ldots$).  A
reasonable set of Pad\'e approximants to $\beta$-functions extracted by
this procedure is shown in \cite{cvetic} to exhibit poles of a magnitude ($0.29\lesssim
x\lesssim 0.33$) sufficiently large for chiral symmetry breaking
at the infrared boundary of QCD.

\smallskip
\noindent
{\bf Acknowledgments:} We are grateful for useful discussions with V.A.\ Miransky, and for support
from the Natural Sciences and Engineering Research Council of Canada (NSERC).

\clearpage

\clearpage

\end{document}